\documentclass[preprint,12pt]{elsarticle}
\usepackage{epsfig}
\usepackage{graphics}
\usepackage{graphicx}
\usepackage{amssymb}
\usepackage{lineno}
\usepackage{xspace}

\journal{Nucl. Instru. Meth. A}

\begin{document}

\def\Journal#1#2#3#4{{#1} {\bf #2}, #3 (#4)}

\def\NCA{Nuovo Cimento}
\def\NIM{Nucl. Instr. Meth.}
\def\NIMA{{Nucl. Instr. Meth.} A}
\def\NPB{{Nucl. Phys.} B}
\def\NPA{{Nucl. Phys.} A}
\def\PLB{{Phys. Lett.}  B}
\def\PRL{Phys. Rev. Lett.}
\def\PRC{{Phys. Rev.} C}
\def\PRD{{Phys. Rev.} D}
\def\ZPC{{Z. Phys.} C}
\def\JPG{{J. Phys.} G}
\def\CPC{Comput. Phys. Commun.}
\def\EPJ{{Eur. Phys. J.} C}
\def\PR{Phys. Rept.}
\def\JHEP{JHEP}
\def\IEEEJSSC{IEEE Journal of Solid-State Circuits}
\def\IEEETNS{IEEE Trans. Nucl. Sci.}
\def\PPNP{Prog. Part. Nucl. Phys.}

\def\pt{p$_{\rm T}$\xspace}
\def\dedx{dE/dx\xspace}
\def\snn{$\sqrt{\rm s_{\rm NN}}$\xspace}
\def\s{$\sqrt{\rm s}$\xspace}

\begin{frontmatter}



\title{The STAR Vertex Position Detector}

\author[1]{W.J.~Llope$^{*,}$}
\author[1]{J.~Zhou}
\author[1]{T.~Nussbaum}
\author[2]{G.W.~Hoffmann}
\author[3]{K.~Asselta}
\author[1]{J.D.~Brandenburg}
\author[1]{J.~Butterworth}
\author[3]{T.~Camarda}
\author[3]{W.~Christie}
\author[4]{H.J.~Crawford}
\author[5]{X.~Dong}
\author[4]{J.~Engelage}
\author[1]{G.~Eppley}
\author[1]{F.~Geurts}
\author[3]{J.~Hammond}
\author[4]{E.~Judd}
\author[1]{D.L.~McDonald}
\author[4]{C.~Perkins}
\author[3]{L.~Ruan}
\author[3]{J.~Scheblein}
\author[2]{J.J.~Schambach}
\author[3]{R.~Soja}
\author[1]{K.~Xin}
\author[6]{C.~Yang}

\address[1]{Rice University, Houston, Texas 77005}
\address[2]{University of Texas, Austin, Texas, 78712}
\address[3]{Brookhaven National Laboratory, Upton, New York 11973}
\address[4]{University of California, Berkeley, California 94720}
\address[5]{Lawrence Berkeley National Laboratory, Berkeley, California 94720}
\address[6]{University of Science \& Technology of China, Hefei 230026, China}
\address{}

\begin{abstract}
The 2$\times$3 channel pseudo Vertex Position Detector (pVPD) in the STAR
experiment at RHIC has been
upgraded to a 2$\times$19 channel detector in the same acceptance, called
the Vertex Position Detector (VPD). This detector is fully integrated
into the STAR trigger system and provides the primary input to the
minimum-bias trigger in Au$+$Au collisions. The information
from the detector is used both in the STAR Level-0 trigger and offline
to measure the location of the primary collision vertex along the beam pipe
and the event ``start time" needed by other fast-timing detectors in STAR. 
The offline timing resolution of single detector channels in full-energy Au$+$Au collisions
is $\sim$100 ps, resulting in a start time resolution of a few tens of
picoseconds and a resolution on the primary vertex location of $\sim$1 cm. 
\end{abstract}

\end{frontmatter}

\vspace*{2mm}
\noindent $^*$ Corresponding author. {\it E-mail address:} llope@rice.edu. 
\vspace*{2mm}



\section{Introduction}\label{sec:intro}

In full-energy, \snn$\!\!\!=$200 GeV, Au$+$Au collisions at the Relativistic
Heavy-Ion Collider (RHIC), pulses of tens to hundreds of photons from
$\pi^0$ decays stream outwards from the primary collision vertex at the speed of light 
and very close to the beam pipe. A detector capable of measuring the arrival 
times of these photons can thus provide important information for
event triggering, especially on the location of the primary vertex
along the beam pipe, and on the time that the event occurred relative
to free-running master clocks used by other fast timing detectors in the experiment. 
In the Solenoidal Tracker at RHIC (STAR) \cite{ref:star}, the first detector 
capable of such measurements that was implemented was the ``pseudo Vertex Position 
Detector" (pVPD) \cite{ref:tofpnim}. This system consisted of two identical 
assemblies of three readout detectors, one on each side of STAR (east and west), 
with each assembly mounted immediately outside the beam pipe. The pVPD worked well 
in full-energy  Au$+$Au collisions \cite{ref:tofpnim}. However, for lighter beams such as 
p$+$p and also Au$+$Au beams at lower beam energies, the efficiency of the 
pVPD for providing the vertex location and the event time degraded due to the
relatively lower multiplicities of very forward prompt particles in such collisions.
To address this, the pVPD was upgraded to increase the readout detector 
channel count on each side of STAR in the same acceptance. This detector
is called the Vertex Position Detector (VPD). 

Unlike the pVPD, the VPD has been fully integrated into the STAR trigger system.
The VPD provides the primary detector input to the STAR minimum bias trigger in A$+$A
collisions. Each VPD assembly measures up to nineteen times in each event. 
These times are thus available, both online and offline, to measure the location of 
the primary vertex along the beam pipe, $Z_{vtx}$, via the equation,
\begin{equation}
Z_{vtx} = c(T_{east} - T_{west})/2,
\label{eq:zvtx}
\end{equation}
where $T_{east}$ and $T_{west}$ are the times from each of the two VPD assemblies and
$c$ is the speed of light. The event ``start time," which is needed by the STAR 
Time-of-Flight (TOF) and Muon Telescope Detector (MTD) \cite{ref:startof} systems 
to perform particle identification at mid-rapidity, is given by,
\begin{equation}
T_{start} = (T_{east} + T_{west})/2 - L/c, 
\label{eq:tstart}
\end{equation}
where $L$ is the distance from either assembly to the center of STAR.

One motivation for the increased channel count from the pVPD to the VPD
in the same angular acceptance was to increase the efficiency of the detector 
for recording hits in single events. Another important motivation involves
the fact that the experimental resolution on either $T_{east}$ or $T_{west}$ used
in equations (\ref{eq:zvtx}) and (\ref{eq:tstart}) improves like 1/$\sqrt{N}$, where
$N$ is the number of channels on each side that were hit by prompt particles. 
That is, taking $T_{east}$ or $T_{west}$ as the average over all channels hit by prompt 
particles in an event results in a resolution on, {\it e.g.}, $Z_{vtx}$, given by,
\begin{equation}
\sigma(Z_{vtx}) = (c/2)\sigma_{\Delta T} = (c/\sqrt{2})\sigma_{T} = (c/\sqrt{2})\sigma_0/\sqrt{N},
\label{eq:zres}
\end{equation}
where $T$ is $T_{east}$ or $T_{west}$, $\sigma_{\Delta T}$ is the resolution of the time
difference $T_{east}$$-$$T_{west}$, $\sigma_T$ is the resolution on $T_{east}$ or $T_{west}$, 
and $\sigma_0$ is the time resolution of a single readout detector. The larger channel count in 
the VPD thus allows an averaging over a larger number of lit channels and hence an improved 
performance for measuring both $Z_{vtx}$ and $T_{start}$. The larger number of readout
channels also provides more time values in the same acceptance, 
allowing one to better recognize which times are prompt and which 
times should be rejected as (delayed) outliers. The rates for such non-prompt
hits are nearly negligible in full-energy Au$+$Au collisions, but are significant
for p$+$p and lower energy Au$+$Au collisions. 

This paper is organized as follows. Section \ref{sec:design} describes
the design of the detectors and the electronics. Section \ref{sec:calibandperform}
describes the configuration of the system for a new RHIC beam, the 
offline calibration of the detectors, and the start-timing and $Z_{vtx}$ performance.
Finally, Section \ref{sec:summary} presents the summary and conclusions.

\section{Design}\label{sec:design}

The VPD exists as two identical detector assemblies, one on the east and
one on the west of STAR.  
The design of these assemblies is described in section \ref{sec:dets}. 
Each of the nineteen detectors used in each assembly is composed of
a Pb converter followed by a fast, plastic scintillator which is read out
by a photomultiplier tube (PMT). The signals
from the nineteen detectors in each assembly are digitized independently
by two different sets of electronics which are described in section
\ref{sec:electronics}.

The detector was first installed in advance of the 2007 RHIC run. The
bases for the PMTs were revised before the 2008 RHIC run. The detector
has been used without additional modifications in every RHIC run since then.

\subsection{Detectors}\label{sec:dets}

\begin{figure}[htbp]
\begin{center}
\includegraphics[keepaspectratio,width=0.7\textwidth]{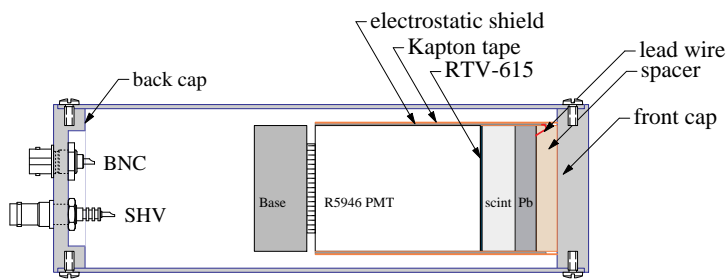}
\vspace*{-4mm}
\caption{A schematic side view of VPD detector. The signal and high voltage
connectors on the back cap are not shown.} 
\label{fig:det}
\end{center}
\end{figure}

Each VPD assembly consists of nineteen detectors, a side view of which is shown
in Figure \ref{fig:det}. Each detector housing is a 2 inch outer diameter and 0.049 inch thick 
aluminum cylinder with 3/8 inch thick aluminum front and back caps. Inside this cylinder
is a 0.25 inch non-conducting spacer, then the active elements consisting of
a 0.25 inch (1.13 radiation lengths) Pb converter, and a 1 cm thick scintillator 
(Eljen EJ-204) coupled to a 1.5 inch diameter Hamamatsu R-5946 mesh dynode PMT 
via RTV-615 optically transparent silicone adhesive. The PMTs used in the VPD 
were taken from the TOFp detector \cite{ref:tofpnim} after it was decommissioned 
in 2005. 

The PMT dynode voltages are provided by a conventional linear resistive base.
An initial version of these bases was used in the 2007 RHIC run, and the
detectors performed well, but there was some ``ringing" in the trailing edge
of the PMT line shapes. This slightly complicated the offline slewing corrections, 
but did not affect the timing performance. In 2008, these bases were revised.
No changes were made to the component values or connections, but the placement
of these components was revised to minimize the lengths of the traces inside
each PMT base circuit board. The excessive trace lengths in the previous
version of the bases resulted in enough distributed inductance to cause
the trailing edge ringing, which was completely removed in the final bases. 

A wire connects the PMT cathode pin to a 0.001 inch thick 
aluminum cylinder which extends past the active elements. Another wire is used to connect 
the aluminum cylinder to the lead converter. In this way, 
the active elements are enclosed on all but one side by an electrostatic shield. 
This shield is electrically isolated from the active elements inside and the aluminum 
outer cylinder outside by several layers of Kapton tape. 
The output coaxial connector shield is isolated from the detector 
housing and the high voltage ground but is indirectly connected via a 
1 k$\Omega$ resistor.  This prevents the (inductive) shield of the 
coaxial signal cable from forming an undesirable resonant circuit with 
the (capacitive) electrostatic shield and detector housing while 
maintaining the high voltage ground return path.

\begin{figure}[htbp]
\begin{center}
\includegraphics[keepaspectratio,width=0.57\textwidth]{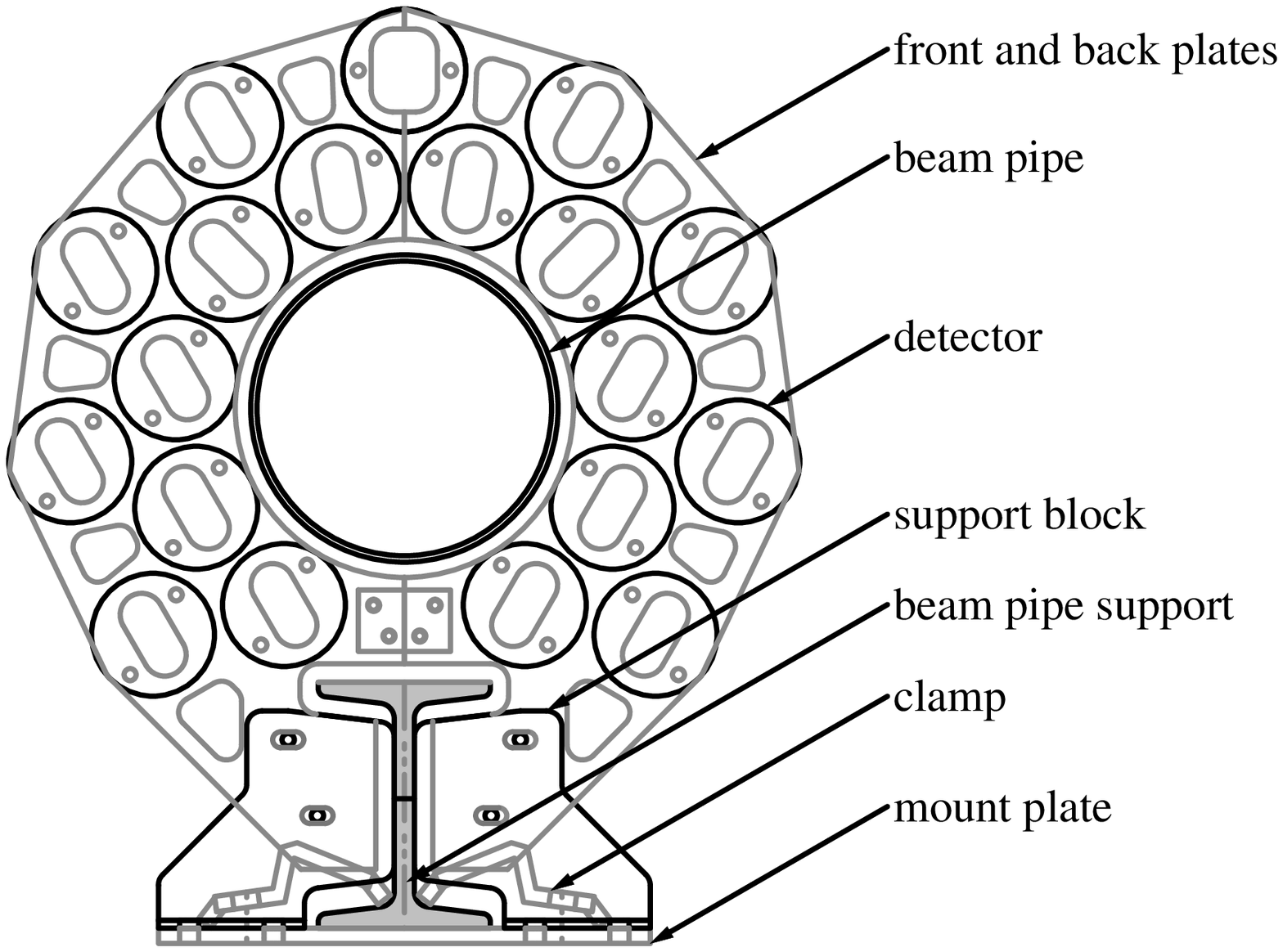}
\includegraphics[keepaspectratio,width=0.41\textwidth]{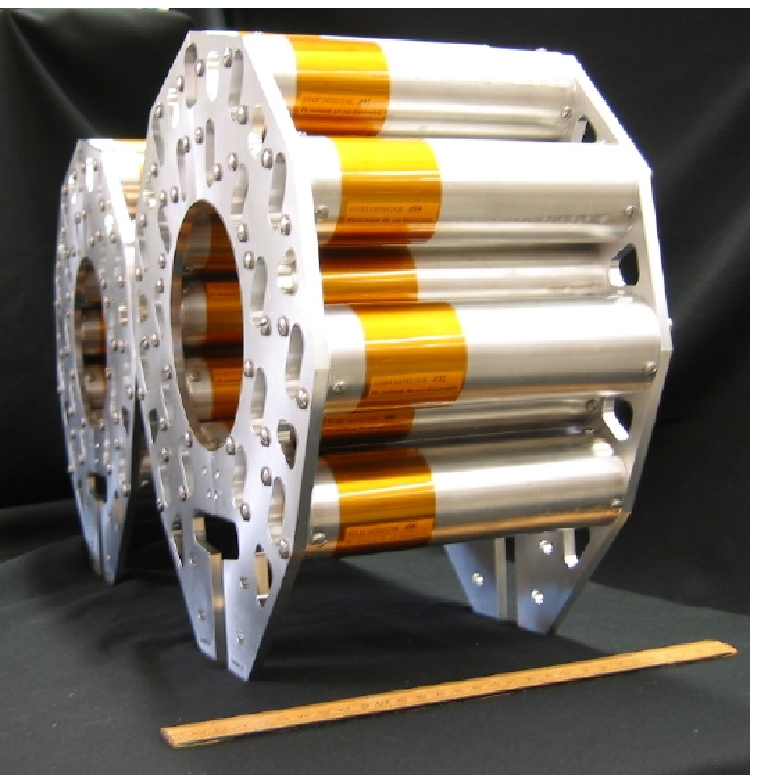}
\vspace*{-4mm}
\caption{On the left is a schematic front view of a VPD assembly, and
on the right is a photograph of the two VPD assemblies. A one foot long ruler
is shown for scale on the right. } 
\label{fig:assy}
\end{center}
\end{figure}

Each VPD assembly consists of two rings of readout detectors and is mounted to the I-beam 
that supports the STAR beam pipe. A front view of one of the VPD 
assemblies is shown in Figure \ref{fig:assy}. The outer diameter of the beam 
pipe at this distance is five inches. An assembly exists as two semi-annular 
``clam-shells" that enclose the beam pipe.
These are bolted together and are held in place by Delrin support blocks which attach
to a horizontal mount plate which is clamped to the beam pipe support I-beam.
The beam pipe and I-beam are at a different (dirty) electrical ground than the
experiment, so the Delrin support blocks both hold the assembly in place and
electrically isolate it. 

The two assemblies are mounted symmetrically with respect to the center of
STAR at a distance of 5.7 m. The nineteen detectors in each assembly
subtend approximately half of the solid angle in the pseudo-rapidity
range of 4.24$\leq$$\eta$$\leq$5.1. When viewed from the rear and looking towards
the center of STAR, the detectors are numbered 1-10(11-19) counter-clockwise
starting from the lower right in Figure \ref{fig:assy} in the inner(outer) ring. 

\subsection{Electronics}\label{sec:electronics}

\begin{figure}[htbp]
\begin{center}
\includegraphics[keepaspectratio,width=0.8\textwidth]{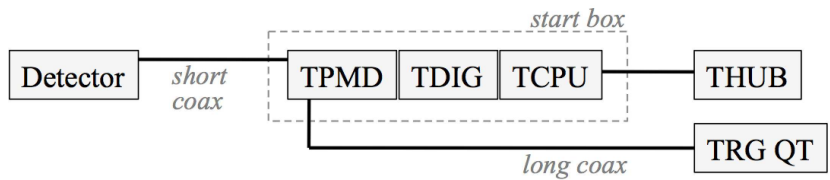}
\vspace*{-4mm}
\caption{A schematic view of the VPD electronics and read-out paths.} 
\label{fig:electronics}
\end{center}
\end{figure}

The signals from the VPD detectors are digitized by two different sets
of electronics. A schematic view of the components that are involved is shown in 
Figure \ref{fig:electronics}.

The signals from the VPD detectors are connected via low-attenuation \cite{ref:tofpnim}
Belden 9310 RG-58 ``short coaxial" cables to ``TPMD" boards which are mounted inside an 
aluminum ``start box" that is positioned within $\sim$6 feet of the VPD detector assemblies. 
Also mounted inside this box are the ``TDIG" and ``TCPU" boards. Digital data are sent over 
Category-6 cables to ``THUB" boards, which then transmit the data over optical fiber to 
receivers in the STAR data acquisition system \cite{ref:daq}. Additional details on these electronics 
boards can be found in Refs. \cite{ref:startof} and \cite{ref:tofelectronics}.

The TPMD boards split the detector signals after they pass a
diode-based clamping circuit which limits the pulse height to 4V. One output 
undergoes a leading-edge discrimination of these signals with respect to a
tunable threshold at typically 25 mV. The Maxim MAX9601 comparator
outputs are sent to the TDIG boards which use the CERN HPTDC
chip \cite{ref:hptdc} to digitize the arrival times of the signals with respect to an
externally-input free-running clock.
Each TDIG board uses three HPTDC chips in ``very-high resolution" mode. 
The digitized times are 21 bit data words with a dynamic range of 52 $\mu$s and
a least significant bit (LSB) time conversion of 25ns/1024 $\approx$ 24.4 ps. The pulse sizes 
are measured by the 
TDIG boards as the width of the pulse at the threshold, which is called the
``Time over Threshold" (ToT). All 2$\times$19 channels of the VPD are digitized
in these electronics. 

The HPTDC chips exhibit a timing cross-talk \cite{ref:xtalk} 
of up to three LSB when two signals arrive at the same HPTDC chip 
within a few nanoseconds of each other. To avoid this, there are five TPMD$+$TDIG
board pairs in each start box, and each board pair handles only 3-4 detector channels
out of the 24 possible. 
When two detector channels are connected to the same HPTDC chip, the ``short coax" 
input cables are staggered in length in 12~ns steps so
as to remove the possibility of HPTDC timing cross-talk.

The other signal from the passive split performed in the TPMD
boards is passed on with negligible attenuation and distortion since
the stub feeding the high impedance comparator is very short.
This signal is sent over long ($\sim$100 ft) coaxial cables to electronics 
in the STAR trigger system called ``QT boards" \cite{ref:qt}.
Short sections of coaxial cable are used
after the TPMD boards and before these ``long coax" cables to equalize the signal
arrival times at the QT electronics to a few nanoseconds given the presence of the
short coax cable staggering before the TPMD boards used to avoid the HPTDC cross-talk.

The QT boards are 9U VME boards and are part of the STAR Trigger system \cite{ref:trg}. They perform
pulse area and time measurements using an analog-to-digital (ADC) and time-to-amplitude-conversion
(TAC) circuitry for all of the STAR trigger detectors. The ADC and TAC measurements
are each 12-bit numbers. The TAC time measurement is common-stop with respect to the
9.4 MHz RHIC clock with a digital-to-time conversion of $\sim$18 ps/LSB. For system design 
simplicity, only 2$\times$16 channels of the VPD are digitized in QT boards.


\section{Calibrations and performance}\label{sec:calibandperform}

The information needed from the VPD for use in equations (\ref{eq:zvtx}) and (\ref{eq:tstart}) 
are time values with the best possible experimental resolution. This first requires that the
high voltage values used for each of the VPD PMTs result in equalized gains, {\it i.e.}~the 
pulses from the different detectors should have the same size for the same amount of 
light generation in the scintillators. 
The time values digitized by the electronics described in section \ref{sec:electronics} 
include offsets resulting from propagation times inside the PMTs, signal cabling, and trace 
lengths in the electronics. There is also a dependence of the digitized pulse times on the size
of the detector pulses which is called ``slewing" \cite{ref:slewing}. The contributions
from these sources of smearing are removed via offline calibrations of the VPD
data. These calibrations, and the gain-matching of the VPD PMTs, is described
in this section. The timing performance of the VPD detectors following these calibrations
will also be presented in this section.

\subsection{Configuration}\label{sec:config}

RHIC is an extremely flexible machine. In the six runs (one per year) since
the VPD was installed, it has provided p$+$p collisions at beam energies, \s,
of 200 and 510 GeV, as well as Cu$+$Au, Au$+$Au, and U$+$U collisions at a number 
of beam energies, \snn, ranging from 7.7 to 200 GeV. The species and the number of particles per event 
producing signals in the VPD detectors depend strongly on the RHIC entrance 
channel and the collision centrality. In Au$+$Au collisions at 200 GeV, the
signal in a VPD detector in one event is produced by 
\raisebox{-0.6ex}{$\stackrel{>}{\sim}$}10 photons from $\pi^0$ decays and
an approximately equal number of charged pions. In the lowest energy Au$+$Au 
collisions, {\it i.e.}~7.7 GeV, a VPD hit results from one or two spectator protons.
The beam energy at which these trends cross depends on the centrality but is 
near 30-50 GeV according to {\sc UrQMD} \cite{ref:urqmd} simulations. In similarity to the full-energy 
Au$+$Au collisions, VPD hits in p$+$p collisions result from a mixture of photons from $\pi^0$'s 
and charged pions, but unlike the full-energy Au$+$Au collisions, a VPD hit 
results from generally only one such particle per event. In order to provide consistently 
useful information from the VPD in these very different beam environments, the gains
of the VPD PMTs must be set for each RHIC beam separately. 

When a new beam is made available to the RHIC experiments, STAR sets up
a simple interaction trigger generally based on the Zero Degree Calorimeter \cite{ref:trg}. 
Using this trigger, data are collected from
the VPD using three different sets of high voltage values, one at best guess 
values and the other two at $\pm$100 V with respect to the best guess values.
The average ADC values from the TRG QT boards, $\langle ADC \rangle$, and the 
average ToT values from the TDIG electronics, $\langle ToT \rangle$, are measured 
for each of the three gain sets.
The dependence of both quantities on the PMT high voltage value is fit
with a power law for each channel. For the 2$\times$16 VPD channels digitized
by the trigger system QT boards, these functions are interpolated to determine the high voltage 
resulting in $\langle ADC \rangle$$=$300. The relationship between the best high voltage
values obtained from the $\langle ADC \rangle$ 
and $\langle ToT \rangle$ values is linear, which allows the gains
for the 2$\times$3 channels not digitized in the QT boards to be consistently 
set on the basis of the $\langle ToT \rangle$ values. 
The high voltage set-points so calculated are $\sim$400 V lower in 200 GeV 
Au$+$Au collisions than they are in 200 GeV p$+$p collisions. The slopes of the
power-law gain curves vary by 6-8\% for the different detectors and are $\sim$5.8(7) 
in 200 GeV p$+$p(Au$+$Au) collisions.

Once the gains are set, data are collected to check the final gains and to set
the basic timing offsets in the QT boards. These offsets are determined by plotting
the time in a VPD channel minus the average over the times in all other lit channels
on the same side of STAR in the same event. The average values of these ``1-$\langle N \rangle$"
distributions are then set as the TAC offset values in the QT boards.

At this point, the VPD is commissioned for the new RHIC beam. Minimum bias triggers
in STAR commonly use the VPD as the primary detector input. These triggers generally
include constraints on the VPD time difference, {\it i.e.}~equation (\ref{eq:zvtx}),
in order to enhance the rate for collisions near the center of STAR which have
the best particle measurement efficiencies and the lowest backgrounds
from particles produced in various detector support materials. These $Z_{vtx}$
constraints are defined via upper and lower limits on the difference between
the uncalibrated times from the earliest VPD hit on each side of STAR. Such
an approach results in a $Z_{vtx}$ resolution of $\sim$5.5 cm in full-energy
Au$+$Au collisions. 

\subsection{Slewing and offset calibrations}\label{sec:calib}

The integrated non-linearity in the HPTDC chips is corrected via 
code-density tests \cite{ref:tofelectronics}.
The smearing of the detector time values resulting from channel-dependent
propagation delays, {\it a.k.a.} offsets, and from the slewing are
removed via offline calibrations. The procedure is iterative, and
results in a table of values binned by the pulse size, ToT, for each channel.
In subsequent analyses, the event-by-event value of the pulse size in
a lit detector in an event is used to look up a value from these tables
which is subtracted from the raw value to remove the offsets and slewing. 
These calibration tables are determined as follows. 

The events used are from the minimum bias trigger, often based on the
VPD data itself, which requires that at least one VPD channel on each side
of STAR is lit in an event. The location of the collision vertex
along the beam pipe as reconstructed from the primary tracks in the
STAR Time Projection Chamber (TPC) \cite{ref:tpc}, $Z_{vtx}^{TPC}$, was required 
to be within $\pm$50 cm of the center of STAR. The transverse radius of the primary 
collision vertex was required to be less than 2 cm (inside the 2.5 cm radius beam pipe). 
At least two primary tracks were also required. 

The first step involves the determination of initial relative offsets. 
These are obtained as the average values of the times in specific lit channels
with respect to a reference channel (channel 1 on the west). Once these offsets are 
obtained, they are applied to the raw time values. The resulting
time values from a VPD channel minus the average of the times of all other
lit channels on the same side of STAR in the same event are plotted as a function
of that channel's ToT value. The dependence
of the average ``1-$\langle$$N$$\rangle$" times on each channel's ToT value
is used in subsequent passes through the data to reduce the slewing and offsets. 
The procedure is iterative and determines three such calibration curves for each channel.
The two VPD assemblies, one east and one west, are treated in parallel. 
The range of the dependence of the 1-$\langle$$N$$\rangle$ time differences on 
the ToT values is generally a few ns before the first pass, a few 
hundred ps before the second pass, and is negligible after the third pass. 
At the end of the procedure, the initial offsets and the three slewing correction tables
for each channel are added together to produce the final correction
tables versus the ToT values. 

In full-energy Au$+$Au collisions, every VPD channel is hit by 10's of prompt
particles in every event. As the beam energy of Au$+$Au collisions is decreased, 
the number of channels lit by prompt particles strongly decreases, and the probability
that VPD channels are lit by non-prompt particles increases. These ``late hit" channels
must be rejected. This outlier rejection is performed
as follows. The location of the primary vertex along the beam pipe calculated from
the VPD times, $Z_{vtx}^{VPD}$, is calculated using equation (\ref{eq:zvtx})
from every possible pair of lit channels with one channel on the east and one channel
on the west. If the $Z_{vtx}^{VPD}$ value from a pair of channels is consistent with 
the value of $Z_{vtx}^{TPC}$, these two channels are flagged as having prompt hits.  
This consistency is with respect to an analysis cut that becomes tighter in each calibration
pass. It is also required that each of the $N$ times used in the 1-$\langle$$N$$\rangle$
relative time calculation are consistent with the average values, $\langle$$N$$\rangle$, 
with a $\sim$4$\sigma$ cut that also becomes tighter in each calibration pass. 

\begin{figure}[htbp]
\begin{center}
\includegraphics[keepaspectratio,width=0.7\textwidth]{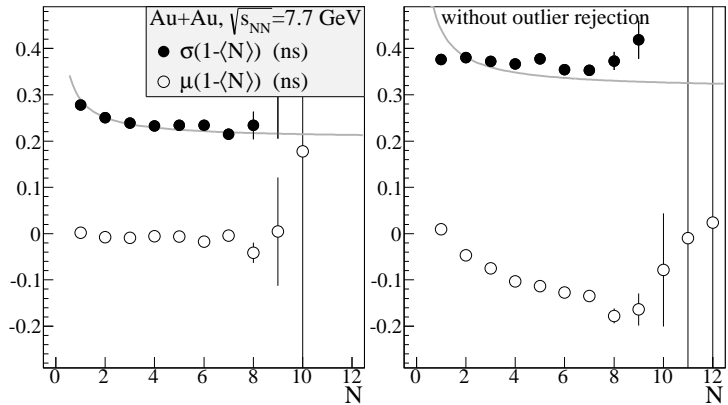}
\vspace*{-4mm}
\caption{The average values (open circles) and standard deviation (solid circles)
of the Gaussian fits to the ``1-$\langle N \rangle$" time differences versus
$N$ for a typical VPD detector channel in the 7.7 GeV Au$+$Au data. The functional
fit (see text) allowing the extraction of the single detector resolution, $\sigma_0$,
is shown as the solid line. In the left frame, the full outlier rejection is performed,
and in the right frame, no outlier rejection is performed.} 
\label{fig:sigma0_meansigm}
\end{center}
\end{figure}  

Following the calibration procedure, the time resolution of a VPD channel after 
this procedure is called the single detector resolution, $\sigma_0$, and is 
extracted as follows. Two dimensional plots of the ``1-$\langle$$N$$\rangle$" time 
differences versus $N$ are filled, and each bin of the $N$ x-axis is fit with a 
Gaussian. The dependence of the standard deviations and mean values from these Gaussian 
fits as a function of $N$ for a typical VPD channel is shown in the left frame of
Figure \ref{fig:sigma0_meansigm}. The solid line in this frame is the fit to 
the standard deviations using the functional form 
$\sigma$(1-$\langle$$N$$\rangle)$$=$$\sigma_0/\sqrt{N/(N+1)}$, where the
single detector resolution, $\sigma_0$, is the fit parameter. 

Important requirements of a successful VPD calibration are that the mean 
values of the ``1-$\langle$$N$$\rangle$" time 
differences (open points) are near zero for all values of $N$, and that the standard
deviations (solid points) follow the expected $\sigma_0/\sqrt{N/(N+1)}$ trend
shown as the solid line in Figure \ref{fig:sigma0_meansigm}. These requirements
indicate the extent to which the VPD times measured by the different detector channels
are self-consistent and result only from prompt particles. The right
frame of Figure \ref{fig:sigma0_meansigm} shows the same standard deviations and mean 
values from the Gaussian fits to the 1-$\langle$$N$$\rangle$ time differences 
as a function of $N$ but when no outlier rejection is performed. In this
frame, the standard deviations (solid points) are $\sim$100~ps larger and 
no longer follow the expected $\sigma_0/\sqrt{N/(N+1)}$ trend, and the mean values 
(open points) deviate significantly from zero. This underscores the importance of
careful VPD time outlier rejection especially in p$+$p and low energy Au$+$Au 
collisions. 

\begin{figure}[htbp]
\begin{center}
\includegraphics[keepaspectratio,width=0.7\textwidth]{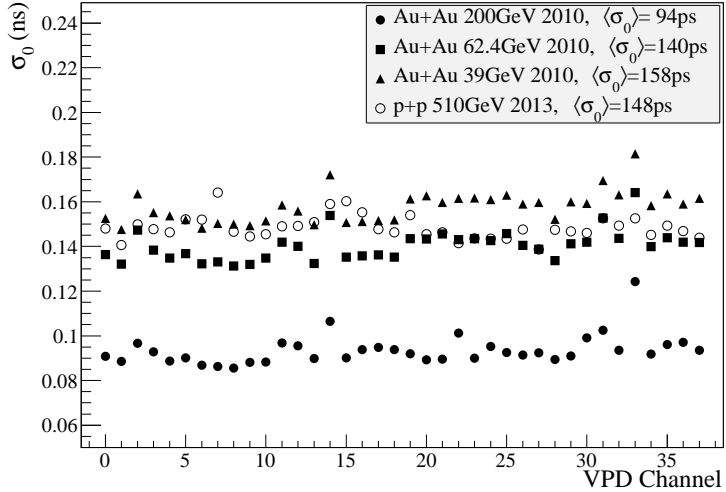}
\vspace*{-4mm}
\caption{The single detector resolution versus the VPD channel number for
a few recent RHIC beams and energies. The results for the VPD detectors on the 
west(east) are shown in the x-axis range 0-18(19-37). } 
\label{fig:sigma0}
\end{center}
\end{figure}  

\subsection{Performance}\label{sec:perform}

The single detector resolution, $\sigma_0$, values extracted as described in the
previous subsection are plotted versus the VPD channel number in Figure \ref{fig:sigma0}
for minimum bias data from a number of different RHIC entrance channels. The
average values of the single detector resolution is 94 ps for full-energy Au$+$Au
collisions, and increases to $\sim$150 ps for intermediate energy Au$+$Au collisions
and 510 GeV p$+$p collisions. The resolution for 7.7 GeV Au$+$Au collisions is approximately
200 ps, and is not shown as the VPD is generally not used to provide the start time
to the TOF system due to its relatively low efficiency per event for those beams. 
The increasing single detector resolution from full-energy Au$+$Au to the lower
energy Au$+$Au and p$+$p collisions reflects the fact that the VPD is doing multiple-particle
timing (per detector channel) in the highest energy Au$+$Au collisions and there is a 
gradual evolution to single-particle timing in p$+$p and lower energy Au$+$Au collisions.

The resolution by which the VPD measures the start time needed by the TOF and MTD
detectors, {\it i.e.}~equation (\ref{eq:tstart}), goes like $\sim$$\sigma_0$/$\sqrt{M}$ where
$M$ is the total number of VPD channels lit by prompt particles in an event. In full-energy Au$+$Au
collisions, the start time resolution is observed to be 20-30 ps, which is essentially
negligible compared to the stop time resolution of the TOF(MTD) detectors of 
$\sim$80(100) ps \cite{ref:startof}. In p$+$p collisions, the VPD start time resolution
is approximately 80 ps, as for those beams $\sigma_0$ is $\sim$150 ps and $M$$\approx$3.

\begin{figure}[htbp]
\begin{center}
\includegraphics[keepaspectratio,width=0.9\textwidth]{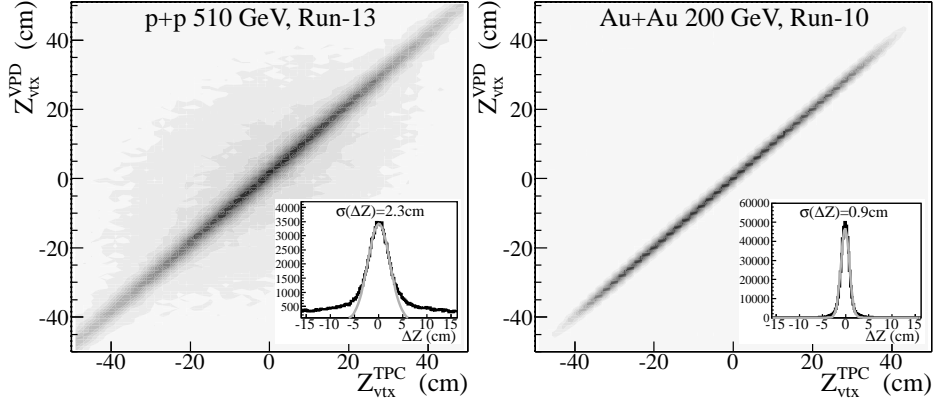}
\vspace*{-4mm}
\caption{The primary vertex position along the beam pipe measured by the VPD, $Z_{vtx}^{VPD}$,
versus the same position as obtained using the primary tracks reconstructed
in the TPC, $Z_{vtx}^{TPC}$ in 510 GeV p$+$p collisions
(left frame) and 200 GeV Au$+$Au collisions (right frame). The insets depict the difference
$\Delta Z$ $=$ $Z_{vtx}^{VPD}$-$Z_{vtx}^{TPC}$ which allows the extraction of the VPD's $Z_{vtx}$ 
resolution as the standard deviation of the difference distributions.} 
\label{fig:zz}
\end{center}
\end{figure}  

The resolution by which the VPD measures the primary vertex location is determined
by comparing the VPD's measurement, $Z_{vtx}^{VPD}$ ({\it i.e.}~equation ({\ref{eq:zvtx})},
to that obtained from the primary tracks reconstructed in the TPC, $Z_{vtx}^{TPC}$.
These two quantities are plotted in Figure \ref{fig:zz} for 510 GeV p$+$p collisions (left frame) 
and 200 GeV Au$+$Au collisions (right frame). 
The insets in each frame depict the difference $\Delta Z$ $=$ $Z_{vtx}^{VPD}$ $-$ $Z_{vtx}^{TPC}$
and Gaussian fits to determine the vertex resolution.
The standard deviations of these distributions obtained from the fits are typically $\sim$2.4 cm 
and $\sim$1 cm in 510 GeV p$+$p collisions and 200 GeV Au$+$Au collisions, respectively.


\section{Summary and conclusions}\label{sec:summary}

The 2$\times$3 channel ``pVPD" \cite{ref:tofpnim} vertex and start-timing detector in the
STAR experiment at RHIC has been replaced by a 2$\times$19 channel detector in the same 
acceptance. This Vertex Position Detector (VPD) exists as two identical assemblies, one on 
each side of STAR, very close to the beam pipe and $\sim$5.7 m from the center of STAR. The 
readout channels in each assembly include a Pb converter followed by a fast plastic scintillator 
and a mesh dynode PMT. The PMT signals are digitized by two different sets of electronics 
for use in the STAR Level-0 trigger to select minimum bias collisions, to constrain 
the location of the primary collision vertex along the beam pipe, and to provide the 
start time needed by other fast timing detectors in STAR.

The system must be configured for each RHIC beam separately to provide a consistent
performance despite the wide range of beam particles (protons to Au) and beam energies
(7.7 to 510 GeV) provided by the RHIC. The slewing and offset corrections are performed 
using an iterative procedure and require a careful rejection of outlier times 
from non-prompt particles. 

The single-detector resolution of the VPD, $\sigma_0$, is approximately 95~ps in full-energy
Au$+$Au collisions, and degrades to $\sim$150~ps in p$+$p and lower energy Au$+$Au collisions.
The start time resolution of the VPD ranges from 20-30 ps in full-energy Au$+$Au collisions
to $\sim$80 ps in p$+$p collisions. The resolution by which the VPD can measure
the location of the primary vertex in full-energy p$+$p and Au$+$Au collisions is 
$\sim$2.5 cm and $\sim$1 cm, respectively. 


\section{Acknowledgments}
We thank the STAR Collaboration for the use of the experimental
data shown in this paper and the operation of this system during RHIC running 
periods as part of STAR standard shift crew operations. 
We thank Allan Schroeder and the members of the UT-Austin machine shop for
the machining of the structural parts of the detector assemblies. 
We appreciate the expert assistance of the BNL
Collider-Accelerator department technicians Charlie Bloxson, Matt Ceglia, and Robbie Karl.
We gratefully acknowledge funding from the US Department of Energy under Grant 
numbers DE-FG02-10ER41666 and DE-FG02-94ER40845.


\end{document}